\documentclass[journal]{IEEEtran}
\IEEEoverridecommandlockouts
\usepackage{amsmath,amssymb,amsfonts}
\usepackage{graphicx}
\usepackage[font=small]{caption}
\usepackage{bm}
\usepackage{color}
\usepackage[colorlinks=false,hidelinks]{hyperref}

\begin{document}

% Copyright
\onecolumn
\thispagestyle{empty}
\Huge IEEE Copyright Notice \\

\large
© 2025 IEEE. Personal use of this material is permitted. Permission from IEEE must be obtained for all other uses, in any current or future media, including reprinting/republishing this material for advertising or promotional purposes, creating new collective works, for resale or redistribution to servers or lists, or reuse of any copyrighted component of this work in other works.

\vfill
This work has been accepted for publication in \textit{IEEE Communications Letters} (Volume: 29, Issue: 11, November 2025). The final published version is available via IEEE Xplore, DOI: \href{https://doi.org/10.1109/LCOMM.2025.3600591}{10.1109/LCOMM.2025.3600591}.

\clearpage
\twocolumn
\normalsize

% Article
\title{\LARGE Design of RIS-aided mMTC+ Networks for Rate Maximization under the Finite Blocklength Regime with Imperfect Channel Knowledge}
\author{Sergi~Liesegang,~\IEEEmembership{Member,~IEEE}, Antonio~Pascual-Iserte,~\IEEEmembership{Senior Member,~IEEE}, \\ Olga~Mu\~noz,~\IEEEmembership{Senior Member,~IEEE}, and Alessio~Zappone,~\IEEEmembership{Fellow,~IEEE}
\thanks{This letter was conducted when S. Liesegang was with the Signal Processing and Communications Group from the Universitat Polit\`ecnica de Catalunya, recognized as a consolidated research group by the Departament de Recerca i Universitats de la Generalitat de Catalunya through 2021 SGR 01033. The work of S. Liesegang, A. Pascual-Iserte, and O. Mu\~noz is part of the I+D+i project 6-SENSES (PID2022-138648OB-I00), funded by MICIU/AEI/10.13039/501100011033 and ERDF/EU. The work of A. Zappone has been funded by the European Union - NextGenerationEU under the project NRRP RESTART, RESearch and innovation on future Telecommunications systems and networks, to make Italy more smART PE\_00000001 - Cascade Call SMART project, with CUP E63C22002040007.}
\thanks{A. Pascual-Iserte and O. Mu\~noz are with the Dept. Signal Theory and Communications, Universitat Polit\`ecnica de Catalunya - BarcelonaTech (UPC), 08034 Barcelona, Spain (e-mails: \{antonio.pascual,olga.munoz\}@upc.edu).}
\thanks{S. Liesegang and A. Zappone are with the Dept. Electrical and Information Engineering, University of Cassino and Southern Lazio (UNICAS), 03043 Cassino, Italy (e-mails: \{sergi.liesegang,alessio.zappone\}@unicas.it).}
}

\ifCLASSOPTIONpeerreview
     \markboth{Accepted for publication in IEEE Communications Letters,~Vol.~29, No.~11, November~2025}
\else
     \markboth{Accepted for publication in IEEE Communications Letters,~Vol.~29, No.~11, November~2025}{Liesegang \MakeLowercase{\textit{et al.}}:  Design of RIS-aided mMTC+ Networks for Rate Maximization under the FBLR with I-CSI}
\fi

\maketitle

\begin{abstract}
Within the context of massive machine-type communications+, reconfigurable intelligent surfaces (RISs) represent a promising technology to boost system performance in scenarios with poor channel conditions. Considering single-antenna sensors transmitting short data packets to a multiple-antenna collector node, we introduce and design an RIS to maximize the weighted sum rate (WSR) of the system working in the finite blocklength regime. Due to the large number of reflecting elements and their passive nature, channel estimation errors may occur. In this letter, we then propose a robust RIS optimization to combat such a detrimental issue. Based on concave bounds and approximations, the nonconvex WSR problem for the RIS response is addressed via successive convex optimization (SCO). Numerical experiments validate the performance and complexity of the SCO solutions.
\end{abstract}


\section{Introduction} \label{sec:1}
\IEEEPARstart{I}{n} the evolution of massive machine-type communications (mMTC), known as mMTC+ \cite{Hua22}, vast sets of simple and low-cost devices autonomously transmit short messages to a base station or collector node (CN). Their large scalability allowed a plethora of unprecedented Internet of Things (IoT) applications in many areas, such as surveillance, smart metering, and remote health monitoring, among others. That makes mMTC+ a key technology for future wireless networks \cite{Ngu22}.

Unfortunately, in many practical scenarios, the retrieval of information can be compromised by unfavorable propagation conditions. To circumvent that, reconfigurable intelligent surfaces (RISs) have been pointed out as an encouraging solution in the last few years \cite{DiR20}. Employing the reflecting elements as phase shifters, these large passive structures are capable of altering the channel to improve the strength of the received signal. More precisely, in mMTC+ deployments with blocking objects, RISs can create alternative paths to reach the CN.

The lack of physical resources and the large dimensionality of mMTC+ systems also represent a significant communication drawback \cite{Dai18}. In particular, to provide simultaneous support to all connected terminals, non-orthogonal transmission becomes mandatory. RISs can then be used to mitigate the impact of high interference and weak channel quality.

Last but foremost, channel state information (CSI) plays a crucial role in harnessing the full potential of systems exploiting RIS \cite{DiR20}. However, feasibly acquiring channel knowledge becomes challenging due to the number of reflecting elements and the absence of processing capabilities at the RIS. As a result, assuming perfect CSI is often unrealistic \cite{Wei21}.

Motivated by our prior work \cite{Lie23}, this letter aims to develop an RIS-assisted scheme that maximizes the weighted sum rate (WSR) in a setup where sensors transmit environmental data. In contrast to \cite{Lie23}, we enhance the performance by equipping the CN with multiple antennas. In that vein, we propose a joint design of the CN spatial filters and the RIS reflection matrix, explicitly accounting for channel estimation errors to ensure a robust solution, an aspect previously neglected in \cite{Lie23}.

Given the use of short packets that is typical in mMTC+ applications (e.g., narrowband IoT), traditional Shannon analysis is no longer valid. Instead, the system must be analyzed under the \textit{finite blocklength regime} (FBLR), where the achievable rate can fall significantly below the channel capacity \cite{Pol10}. This deviation can be captured through a correction term known as \textit{channel dispersion}. Unfortunately, optimizing WSR in the FBLR context results in nonconvex problems. To tackle this, we adopt a successive convex optimization (SCO) approach \cite{Mat20}, replacing certain nonconvex parts with surrogate and approximate functions to derive suboptimal yet practically viable solutions, with theoretical optimality claims.
 
With the above considerations, this work constitutes an extension of \cite{Lie23}, where the same authors assumed perfect CSI and single-antenna receivers. To the best of our knowledge, no other studies have been reported in the same direction so far. For instance, the WSR was also maximized in \cite{Zhi23}, considering imperfect channel knowledge but infinite packets. A similar approach with only partial CSI can be found in \cite{You21}, where the authors derived closed-form expressions to optimize energy efficiency. In \cite{Sol24}, the spectral and energy efficiencies of ultra-reliable low-latency systems were maximized under perfect CSI conditions. Contrarily, partial CSI and FBLR were included in \cite{Elw23}, yet the focus laid on power control.

The remainder of this letter is structured as follows. Section~\ref{sec:2} describes the system model and the optimization problem. Section~\ref{sec:3} presents the proposed solutions. Section~\ref{sec:4} provides the numerical results. Section~\ref{sec:5} concludes the work. 

\section{System Model and Problem Formulation} \label{sec:2}
\begin{figure}[t]
\centering
\includegraphics[trim={0 15 0 0}, clip=true, scale = 0.3]{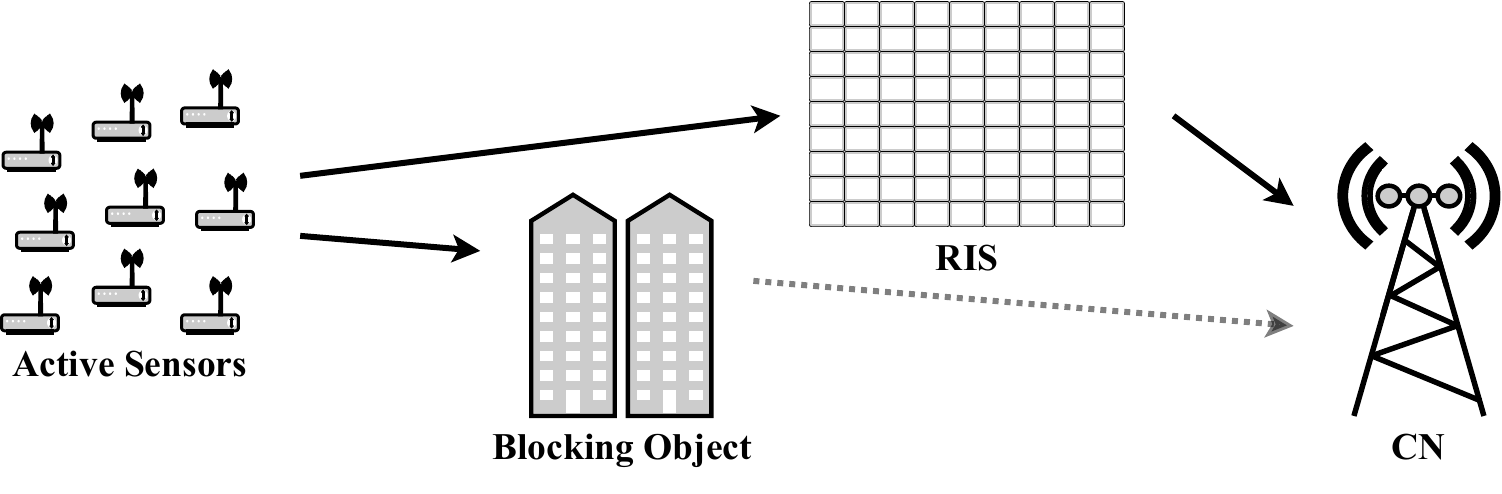}
\caption{Setup with $M = 9$ active sensors, $L = 81$ RIS elements, and $K = 3$ CN antennas. Solid/dotted lines indicate strong/weak paths.}
\label{fig:1}
\end{figure}

This letter will focus on a scenario where $M$ single-antenna sensors send their symbols $x_i \sim \mathcal{CN}(0, P_i)$ to a CN equipped with $K$ antennas. To strengthen communication, we introduce an RIS with $L$ elements (cf. \cite{Lie23}). An illustrative example is shown in Fig.~\ref{fig:1}, where an object blocks the line of sight (LoS) and the RIS helps to spatially focus the signals toward the CN.

The received signal at the CN can be written as
\begin{equation}
    \mathbf{y} \triangleq \sum\nolimits_{i = 1}^M (\mathbf{q}_i + \mathbf{G}\bm{\Psi}\mathbf{g}_i) x_i + \mathbf{w} \in \mathbb{C}^K,
    \label{eq:1}
\end{equation}
where $\mathbf{q}_i \in \mathbb{C}^K$ is the (direct) channel between sensor $i$ and the CN, $\mathbf{G} \in \mathbb{C}^{K \times L}$ is the channel between the RIS and the CN, $\mathbf{g}_i \in \mathbb{C}^L$ is the channel between sensor $i$ and the RIS, $\bm{\Psi} \triangleq \textrm{diag}(\lambda_1 e^{j\phi_1},\ldots,\lambda_L e^{j\phi_L})$ is the reflection matrix\footnote{In this work, we assume an RIS modeling without hardware impairments. However, as shown in \cite{Li25,Li24}, such imperfections can severely compromise the system's performance and will be examined in future research lines.} with amplitude coefficients $\lambda_l \in [0,1]$ and phase shifts $\phi_l \in [0,2\pi)$ \cite{Li21}, and $\mathbf{w} \sim \mathcal{CN}(\mathbf{0}_K,\sigma_w^2 \mathbf{I}_K)$ is the thermal noise. 

Given that the direct link might be blocked, we consider $\mathbf{q}_i$ to be negligible compared to the RIS channels $\mathbf{G}$ and $\mathbf{g}_i$ \cite{Lie23}. This means the RIS is imperative for feasible communication. In that sense, motivated by the reasoning in \cite{Zhi23}, we adopt a Rician fading model for the concatenated channels, i.e.,
\begin{align}
    \label{eq:2} [\mathbf{G}]_{k,l} &= \sqrt{\alpha/(1 + \gamma)}([\bar{\mathbf{G}}]_{k,l} + \sqrt{\gamma} [\ddot{\mathbf{G}}]_{k,l}), \\
    \label{eq:3} [\mathbf{g}_i]_l &= \sqrt{\beta_i/(1 + \delta_i)}([\bar{\mathbf{g}}_i]_l + \sqrt{\delta_i} [\ddot{\mathbf{g}}_i]_l),
\end{align}
defined in a general way such that it is valid for both the near- and far-field regions \cite[Subsection II-A]{Liu23}: $\alpha$ and $\beta_i$ are the large-scale fading coefficients that include the propagation losses; $\gamma$ and $\delta_i$ are the Rician factors; $[\bar{\mathbf{G}}]_{k,l} \sim \mathcal{CN}(0,1)$ and $[\bar{\mathbf{g}}_i]_l \sim \mathcal{CN}(0,1)$ are the independent Rayleigh non-LoS paths; while $[\ddot{\mathbf{G}}]_{k,l}$ ($[\ddot{\mathbf{g}}_i]_l$) are the LoS components that depend on the position of the $k$-th CN antenna (sensor $i$) and the $l$-th RIS element. All variables are specified in Section~\ref{sec:4}.

Under the assumption of perfect CSI at the CN, the received signal-to-interference-plus-noise ratio (SINR) when detecting sensor $i$ through a linear spatial filter $\mathbf{f}_i \in \mathbb{C}^K$ yields
\begin{equation}
    \rho_i \triangleq P_i \vert \mathbf{f}_i^{\textrm{H}}\mathbf{H}_i \bm{\psi} \vert^2 / \left(\sigma_w^2 \| \mathbf{f}_i \|_2^2 + \sum\nolimits_{j \neq i} P_j \vert \mathbf{f}_i^{\textrm{H}}\mathbf{H}_j \bm{\psi} \vert^2\right),
    \label{eq:4}
\end{equation} 
where $\mathbf{H}_i \triangleq \mathbf{G} \textrm{diag}(\mathbf{g}_i) \in \mathbb{C}^{K \times L}$ is the cascaded channel and $\bm{\psi} = [\psi_1,\ldots,\psi_L]^{\textrm{T}}$ is the diagonal of the RIS matrix, i.e., $\bm{\Psi} = \textrm{diag}(\bm{\psi})$, with $\psi_l \triangleq \lambda_l e^{j \phi_l}$ such that $\vert \psi_l \vert \leq 1$.

Based on that, the instantaneous data rate of sensor $i$ under the FBLR (in bits/Hz) can be approximated by \cite{Pol10}
\begin{equation}
    R(\rho_i) \approx C(\rho_i) - a_i D(\rho_i),
    \label{eq:5}
\end{equation}
where $C(x) \triangleq \log_2 (1 + x)$ is the channel capacity; $D(x) \triangleq \sqrt{V(x)}$ is the FBLR penalty, with $V(x) \triangleq 2 x(1 + x)^{-1}$ the channel dispersion \cite{Sca17}; and $a_i \triangleq (\log_2 e / \sqrt{n_i}) Q^{-1}(\epsilon_i)$ is a scaling (constant) factor, with $Q(\cdot)$ the Gaussian Q-function, $\epsilon_i$ the error probability, and $n_i$ the number of transmit symbols.

As discussed in \cite{Neu18}, the extension of \eqref{eq:5} to the case of imperfect CSI (I-CSI) is challenging. We thus opt for the lower bound from \cite{Has03} and condition the SINR in \eqref{eq:4} on the channel observations to deal with tractable analytic expressions.

Although the CN can estimate the cascaded channels, a lot of feedback is required to provide this information to the RIS and frequently adapt its response to the temporal variations of $\mathbf{H}_i$. That is why we will advocate for a more practical two-timescale scheme for jointly designing the CN spatial filters and the RIS reflection matrix (cf. \cite{Zhi23,Pal23}). In short, the set $\{\mathbf{f}_i\}$ will be constructed using instantaneous CSI, whereas the vector $\bm{\psi}$ will be configured based only on channel distribution information, which varies more slowly than CSI \cite{Elw23, Neu18}.

When considering the low-complexity maximum ratio combining (MRC) at reception\footnote{MRC represents a worst-case scenario, as interference is not suppressed. More complex techniques, such as zero-forcing (ZF) \cite{Elw23} or minimum mean-square error (MMSE) \cite{Neu18} filtering, could further enhance performance. Due to lack of space, such extensions will be addressed in future investigations.}, i.e., $\mathbf{f}_i = \hat{\mathbf{H}}_i \bm{\psi}$, we have \cite{You21}
\begin{equation}
    \rho_i = \frac{P_i \vert \bm{\psi}^{\textrm{H}} \hat{\mathbf{H}}_i^{\textrm{H}} \hat{\mathbf{H}}_i \bm{\psi} \vert^2}{\sigma_w^2 \| \hat{\mathbf{H}}_i \bm{\psi} \|_2^2 + \|\bm{\psi} \|_2^2\bm{\psi}^{\textrm{H}} \bm{\Lambda}_i \bm{\psi} + \sum\nolimits_{j \neq i} P_j  \vert \bm{\psi}^{\textrm{H}} \hat{\mathbf{H}}_i^{\textrm{H}} \hat{\mathbf{H}}_j \bm{\psi} \vert^2},
    \label{eq:6}
\end{equation}
with $\bm{\Lambda}_i \triangleq \hat{\mathbf{H}}_i^{\textrm{H}} \tilde{\mathbf{C}} \hat{\mathbf{H}}_i$ including the channel estimate $\hat{\mathbf{H}}_i \in \mathbb{C}^{K \times L}$ and the sum of the covariances of the estimation errors $\tilde{\mathbf{C}} \in \mathbb{C}^{K \times K}$. For more details, please see the Appendix.

As a result, since the design variable $\bm{\psi}$ determines the SINR $\rho_i$ and, therefore, the rate in \eqref{eq:5}, i.e., $R_i(\bm{\psi}) \equiv R(\rho_i(\bm{\psi}))$, the maximization of the (ergodic) WSR can be formulated as \cite{Lie23}
\begin{equation}
\max_{\bm{\psi}} \sum\nolimits_{i=1}^M \omega_i \mathbb{E}[C_i(\bm{\psi}) - a_i D_i(\bm{\psi})] \; \; \; \textrm{s.t.} \; \; \; \vert \psi_l \vert \leq 1 \, \forall l,
\label{eq:7}
\end{equation}
where the weights $\omega_i$ establish priorities \cite{Guo20} and the expectation is taken with respect to (w.r.t.) the channel observations. In this work, we focus on optimizing the instantaneous rate and compute such averaging via Monte-Carlo\footnote{Numerical evaluation methods can be computationally expensive, and an asymptotic deterministic equivalent could help to reduce complexity \cite{You21}. However, this is beyond the scope of this letter and is left for future studies.} \cite{Neu18,Zap14}. This leads to tighter expressions than the popular \textit{hardening} bound that assumes knowledge of the effective channel means. It is used in \cite{Zhi23}, where smaller search spaces ($\vert \psi_l \vert = 1$ $\forall l$) allowed the authors to derive analytic closed-form expressions. 

Unfortunately, the problem is not convex w.r.t. $\bm{\psi}$ due to the nonconcavity of the objective function. The sequel is devoted to discussing suboptimal approaches based on concave lower bounds for $R_i$ (also applicable to $\mathbb{E}[R_i]$ since the expectation is a linear operator that preserves convexity/concavity \cite{Boy04}). 

\section{Proposed Solution} \label{sec:3}
In this section, we propose strategies relying on SCO to solve the problem defined in \eqref{eq:7} where, at the $k$-th iteration, we update the solution $\bm{\psi} \equiv \bm{\psi}^{(k)}$ from the previous feasible point, represented by $\bm{\psi}^{(k-1)}$, until convergence is reached.

Following the discussion in \cite[Subsection~III.B]{Lie23}, rather than seeking $\bm{\psi} $ directly, we first introduce $\bm{\Phi} \triangleq \bm{\psi} \bm{\psi}^{\textrm{H}} \in \mathbb{C}^{L \times L}$ (that has rank $1$) as the new optimization variable and later recover the original solution with Gaussian randomization \cite{Wu19}. Note that the previous solution is denoted by $\bm{\Phi}^{(k-1)}$. This iterative procedure therefore yields a suboptimal but feasible $\bm{\psi}$.

Accordingly, the original problem in \eqref{eq:7} is decomposed into a sequence of subproblems that are solved iteratively, i.e.,
\begin{equation}
\begin{aligned}
\max_{\bm{\Phi}} \, \, &\sum\nolimits_{i=1}^M \omega_i \mathbb{E}[C_i(\bm{\Phi}) - a_i D_i(\bm{\Phi})]  \\
\textrm{s.t.} \quad &\bm{\Phi} \succeq \mathbf{0},  \quad \textrm{rank}(\bm{\Phi}) = 1,  \quad  [\bm{\Phi}]_{l,l} \leq 1 \, \forall l,
\end{aligned}
\label{eq:8}
\end{equation}
where $C_i(\bm{\Phi}) \equiv C_i(\bm{\Phi}^{(k)})$ and $D_i(\bm{\Phi}) \equiv D_i(\bm{\Phi}^{(k)})$ denote the capacity and penalty at the $k$-th iteration, respectively.

Among other mild assumptions, for the sequential procedure in \eqref{eq:8} to converge to a stationary point of \eqref{eq:7}, each subproblem must be solved globally \cite[Section IV]{Mat20}. That is why we will seek concave surrogate functions that approximate the original objective function and allow us to find such global optima.

For the sake of clarity, we start with the case of a single antenna at the CN, which extends the main contribution of \cite{Lie23} presented by the same authors. Later, we generalize it to the multiple-antenna setup, i.e., $K > 1$.

\subsection{Single-antenna CN} \label{sec:3.1}
For $K = 1$, the SINR in \eqref{eq:6} can be rewritten as
\begin{equation}
\rho_i(\bm{\Phi}) =\frac{P_i \textrm{tr}(\hat{\mathbf{h}}_i^{\textrm{H}} \hat{\mathbf{h}}_i \bm{\Phi})}{\sigma_w^2 + \tilde{c}\textrm{tr}(\bm{\Phi}) + \sum\nolimits_{j \neq i} P_j \textrm{tr}(\hat{\mathbf{h}}_j^{\textrm{H}} \hat{\mathbf{h}}_j \bm{\Phi})},
\label{eq:9}
\end{equation}
where $\hat{\mathbf{h}}_i \in \mathbb{C}^{1 \times L}$ is the corresponding cascaded channel and $\tilde{c}$ is the sum of (scalar) variances of the estimation errors that can be obtained by particularizing $\tilde{\mathbf{C}}$ for $K = 1$ (cf. Appendix).

In line with \cite{Lie23, Elw23}, here we are also interested in finding a concave lower bound for $C_i(\bm{\Phi})$ and a convex upper bound for $D_i(\bm{\Phi})$. To that end, we consider the inequalities
\begin{gather}
    \label{eq:10} \ln \left(1 + \frac{x}{y}\right) \geq \ln \left(1 + \frac{\bar{x}}{\bar{y}}\right) + \frac{\bar{x}}{\bar{y}} \left(2\sqrt{\frac{x}{\bar{x}}} - \frac{x + y}{\bar{x} + \bar{y}} - 1\right), \\
    \label{eq:11} \sqrt{x} \leq 0.5 (\sqrt{\bar{x}} + x/\sqrt{\bar{x}}),
    \quad x \leq 0.5 (\bar{x} + x^2/\bar{x}),
\end{gather}
for any $x > 0$, $y > 0$, $\bar{x} > 0$, and $\bar{y} > 0$.

Similar to expressions (19) and (21) from \cite{Lie23}, by defining $\bm{\Pi}_i \triangleq P_i \hat{\mathbf{h}}_i^{\textrm{H}} \hat{\mathbf{h}}_i$, $\tilde{\bm{\Pi}}_i \triangleq \tilde{c} \mathbf{I}_L + \sum\nolimits_{j \neq i} P_j \hat{\mathbf{h}}_j^{\textrm{H}} \hat{\mathbf{h}}_j$, and $\bar{\bm{\Pi}}_i \triangleq \bm{\Pi}_i + \tilde{\bm{\Pi}}_i $, we can derive the following concave and convex bounds:
\begin{align}
    \label{eq:12}C_i(\bm{\Phi}) &= \log_2 (1 + \textrm{tr}(\bm{\Pi}_i\bm{\Phi})/(\sigma_w^2 + \textrm{tr}(\tilde{\bm{\Pi}}_i\bm{\Phi})))\\ 
    &\geq C_i(\bm{\Phi}^{(k-1)}) + \frac{1}{\ln 2} \frac{\textrm{tr}(\bm{\Pi}_i \bm{\Phi}^{(k-1)})(\Gamma_i(\bm{\Phi}) - 1)}{\sigma_w^2 + \textrm{tr}(\tilde{\bm{\Pi}}_i \bm{\Phi}^{(k-1)})}, \nonumber \\
    \label{eq:13} D_i(\bm{\Phi}) &= \sqrt{2 \textrm{tr}(\bm{\Pi}_i \bm{\Phi})/(\sigma_w^2 + \textrm{tr}(\bar{\bm{\Pi}}_i\bm{\Phi}))}
    \\ &\leq 0.5 D_i(\bm{\Phi}^{(k-1)}) + 1/(D_i(\bm{\Phi}^{(k-1)})(\sigma_w^2 + \textrm{tr}(\bar{\bm{\Pi}}_i\bm{\Phi}))) \nonumber \\
    &\quad \times 0.5\left(\textrm{tr}(\bm{\Pi}_i \bm{\Phi}^{(k -1)}) + \textrm{tr}^2(\bm{\Pi}_i\bm{\Phi})/\textrm{tr}(\bm{\Pi}_i \bm{\Phi}^{(k -1)})\right), \nonumber    
\end{align}
with
\begin{align}
    \Gamma_i(\bm{\Phi}) 
    \triangleq 2\sqrt{\frac{\textrm{tr}(\bm{\Pi}_i\bm{\Phi})}{\textrm{tr}(\bm{\Pi}_i\bm{\Phi}^{(k-1)})}}  - \frac{\sigma_w^2 + \textrm{tr}(\bar{\bm{\Pi}}_i\bm{\Phi})}{\sigma_w^2 + \textrm{tr}(\bar{\bm{\Pi}}_i \bm{\Phi}^{(k-1)})}.
    \label{eq:14}
\end{align}

Lastly, after applying semi-definite relaxation (SDR) and removing the nonconvex rank-one constraint, we can find the global solution $\bm{\Phi}$ for subproblem \eqref{eq:8} through standard numerical methods (e.g., CVX \cite{CVX20}), and iterate it until converging to a stationary point, i.e., SCO (cf. \cite{Mat20}). However, recall that the final solution might not be the global optimum of \eqref{eq:7}.

\subsection{Multiple-antenna CN} \label{sec:3.2}
Different from \eqref{eq:9}, now the SINR yields
\begin{equation}
    \rho_i(\bm{\Phi}) = S_i(\bm{\Phi})/(W_i(\bm{\Phi}) + E_i(\bm{\Phi}) + I_i(\bm{\Phi})),
    \label{eq:15}
\end{equation}
where, by defining $\bm{\Xi}_{i,j} \triangleq \hat{\mathbf{H}}_i^{\textrm{H}} \hat{\mathbf{H}}_j \in \mathbb{C}^{L \times L}$, we have
\begin{equation}
\begin{array}{ll}
S_i(\bm{\Phi}) \triangleq P_i \textrm{tr}^2(\bm{\Xi}_{i,i} \bm{\Phi}), &  W_i(\bm{\Phi}) \triangleq \sigma_w^2\textrm{tr}(\bm{\Xi}_{i,i} \bm{\Phi}), \\
E_i(\bm{\Phi}) \triangleq \textrm{tr}(\bm{\Phi}\bm{\Lambda}_i \bm{\Phi}), &
I_i(\bm{\Phi}) \triangleq \sum_{j \neq i} P_j \vert \textrm{tr}(\bm{\Xi}_{i,j} \bm{\Phi}) \vert^2.
\end{array}
\label{eq:16}
\end{equation}

Again, by means of the inequality \eqref{eq:10}, we find the bound
\begin{align}
C_i(\bm{\Phi}) \geq C_i(\bm{\Phi}^{(k-1)}) + \rho_i(\bm{\Phi}^{(k-1)}) (\Gamma_i(\bm{\Phi}) - 1)/\ln 2,
    \label{eq:17}
\end{align}
where, letting $T_i(\bm{\Phi}) \triangleq S_i(\bm{\Phi}) + W_i(\bm{\Phi}) + E_i(\bm{\Phi}) + I_i(\bm{\Phi})$,
\begin{align}
    \Gamma_i (\bm{\Phi}) = 2\frac{\textrm{tr}(\bm{\Xi}_{i,i} \bm{\Phi})}{\textrm{tr}(\bm{\Xi}_{i,i} \bm{\Phi}^{(k-1)})} - \frac{T_i(\bm{\Phi})}{T_i(\bm{\Phi}^{(k-1)})}.
    \label{eq:18}
\end{align}

Note that $T_i(\bm{\Phi})$ is convex w.r.t. $\bm{\Phi}$ since $S_i(\bm{\Phi})$ is quadratic, $W_i(\bm{\Phi})$ is linear, $E_i(\bm{\Phi})$ is convex, and $I_i(\bm{\Phi})$ is the sum of quadratic functions. Hence, we end up with a concave lower bound for $C_i(\bm{\Phi})$, from now on denoted by $\tilde{C}_i(\bm{\Phi})$.

To find a surrogate convex upper bound for the correction term $D_i(\bm{\Phi}) = \sqrt{2S_i(\bm{\Phi})/T_i(\bm{\Phi})}$, we start with \eqref{eq:11}:
\begin{equation}
    D_i(\bm{\Phi}) \leq 0.5D_i(\bm{\Phi}^{(k-1)}) + S_i(\bm{\Phi})/(D_i(\bm{\Phi}^{(k-1)})T_i(\bm{\Phi})).
    \label{eq:19}
\end{equation}

Then, by linearizing $T_i(\bm{\Phi})$ via the first-order Taylor series expansions at the previous point \cite{Scu14}, i.e.,
\begin{equation}
    \tilde{T}_i(\bm{\Phi}) \triangleq T_i(\bm{\Phi}^{(k-1)}) + \langle \Delta \bm{\Phi}, \nabla T_i(\bm{\Phi}^{(k-1)}) \rangle,
    \label{eq:20}
\end{equation}
where $\Delta \bm{\Phi} \triangleq \bm{\Phi} - \bm{\Phi}^{(k - 1)}$, $\langle \mathbf{X}, \mathbf{Y}\rangle \triangleq \Re\{\textrm{tr}(\mathbf{X}^{\textrm{H}}\mathbf{Y})\}$ is the Frobenius inner norm for any complex matrices $\mathbf{X},\mathbf{Y}$, and
\begin{align}    
    \nabla T_i(\bm{\Phi}) &=  2 P_i \textrm{tr}(\bm{\Xi}_{i,i} \bm{\Phi})\bm{\Xi}_{i,i}^{\textrm{H}} + \sigma_w^2\bm{\Xi}_{i,i}^{\textrm{H}} + (\bm{\Phi}\bm{\Lambda}_i + \bm{\Lambda}_i\bm{\Phi})^{\textrm{H}} \nonumber \\
    &\quad + \sum\nolimits_{j \neq i} 2P_j \textrm{tr}(\bm{\Xi}_{i,j} \bm{\Phi})\bm{\Xi}_{i,j}^{\textrm{H}},
    \label{eq:21}
\end{align}
is the conjugate gradient, the ratio in \eqref{eq:19} can be safely upper bounded by a quadratic-over-linear function \cite{Sun17}; thus, convex.

Consequently, together with $\tilde{C}_i(\bm{\Phi})$ defined in \eqref{eq:17}, we can obtain a concave lower bound for the FBLR data rate so that problem \eqref{eq:8} can be solved as in Subsection~\ref{sec:3.1}. 
Briefly, (i) drop the rank-one constraint (that is, the SDR); (ii) globally solve the sequence of convex problems through semi-definite programming \cite{Boy04} at each iteration \cite{Mat20} until convergence (that is, the SCO procedure); and (iii) calculate the corresponding final value for $\bm{\psi}$ via Gaussian randomization \cite{Wu19}. 

Recall that, in our previous work \cite{Lie23}, we assumed perfect channel knowledge and $K = 1$. In these subsections, we have presented an RIS-aided system design that is robust to I-CSI errors and formulated the optimization problem for the case of single-/multiple-antenna CN. Because of that, the techniques used in \cite{Lie23} no longer apply, and the bounds for the channel capacity $C_i(\bm{\Phi})$ and FBLR penalty $D_i(\bm{\Phi})$ have been revisited. The above derivations thus characterize a more general setting.

\section{Numerical Evaluation} \label{sec:4}
To assess our approach, we present the WSR w.r.t. the number of CN antennas $K$ and the number of RIS elements $L$. For a broader comparison, we include the random configuration of the phase-shifts $\phi_l$ (denoted by RO), and the case where the coefficients $\psi_l$ are found sequentially using one-dimensional exhaustive searches, i.e., AO (alternating optimization) \cite{Wu19}. In line with the findings in \cite{Lie23}, we will omit gradient descent algorithms, as they often fail in these frameworks.

We consider the indoor-hotspot scenario described in \cite{3GPP36814} with $P_i = 10$ dBm, $\sigma_w^2 = N_o B$, $N_o = -174$ dBm/Hz, $B = 20$ MHz, and a carrier frequency of $2$ GHz. We assume $M = 10$ sensors uniformly distributed within the deployment area of $10$ m\textsuperscript{2} so that the near-field model applies. Note that, given the sporadic nature of mMTC+ applications \cite{Hua22}, $M$ is the number of devices transmitting simultaneously, which is significantly smaller than the total number of deployed devices, say $S$. For example, for periodic reports of $l = 10$ kb every $t = 6$ h at $R = 50$ kbps, supporting $M = 10$ active sensors translates to $S = (t/l) M R \approx 10^6 $ total terminals. The CN and RIS are arbitrarily located at adjacent sides, e.g., ($5$ m, $0$ m) and ($0$ m, $5$ m). We generate the CN and RIS steering vectors according to a uniform linear array and a uniform planar array, respectively \cite{Zhi23}. We also consider $n_i = 100$ transmit symbols (cf. narrowband IoT) and a target PER of $\epsilon_i = 10^{-3}$. 

\subsection{WSR Performance}
\begin{figure}[t]
    \centering    
    \includegraphics[scale = 0.85]{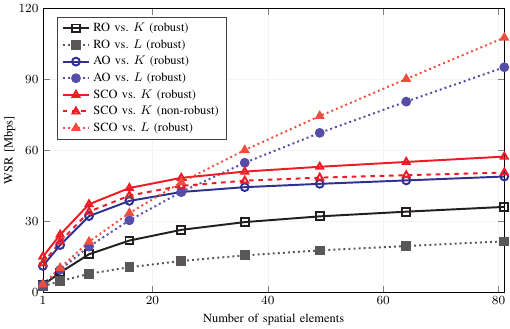}
    \caption{WSR ($\omega_i = 1$ $\forall i$) w.r.t. the number of spatial elements: the number of CN antennas $K$ (solid) and RIS elements $L$ (dotted). \newline}
    \label{fig:2}   
    \centering    
    \includegraphics[scale = 0.85]{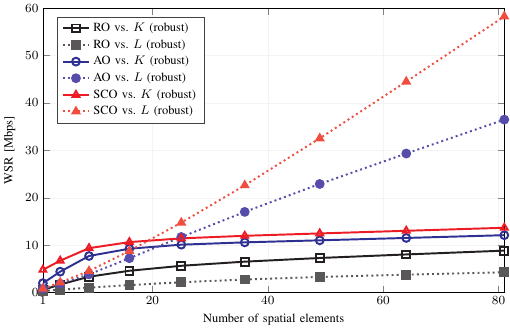}
    \caption{WSR ($\omega_i \propto 1/\rho_i$) w.r.t. the number of spatial elements: the number of CN antennas $K$ (solid) and RIS elements $L$ (dotted). \newline}
    \label{fig:3}
    \centering    
    \includegraphics[scale = 0.85]{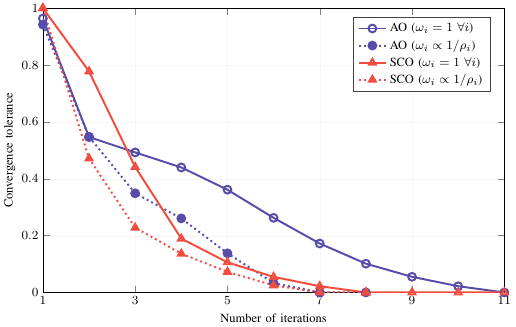}
    \caption{Convergence tolerance w.r.t. the number of iterations.}
    \label{fig:4}    
\end{figure}

The WSR is illustrated in Fig.~\ref{fig:2} and Fig.~\ref{fig:3} for different weight criteria: equality ($\omega_i = 1$ $\forall i$) and fairness ($\omega_i \propto 1/\rho_i$), respectively. The evaluation is done w.r.t. the number of spatial elements (CN antennas $K$ and RIS size $L$). Unless otherwise stated, the baseline configuration is $L = 16$ and $K = 4$.

As pictured in Fig.~\ref{fig:2}, the WSR with equal weights ($\omega_i = 1$ $\forall i$) improves with the spatial degrees of freedom (DoF) in the RIS-CN link. This is expected as more DoF ($K$ or $L$) bring higher beamforming gains. Remarkably, better performance is achieved for high $L$ compared to large $K$. This is because the RIS matrix is optimized, whereas the CN filters remain fixed (MRC). Hence, under this setting, since increasing the number of passive elements is more cost-effective and energy-efficient (unlike the RIS, the power consumption of the CN scales with the size of the array), RIS emerges as a promising solution for green MTC+ systems. In other words, one can consider using smaller arrays in exchange for bigger surfaces with adequately configured reflecting elements. Lastly, we include the solution assuming that the available CSI is perfect, despite not being true (i.e., non-robust design), to illustrate the loss due to the channel knowledge mismatch. This is done only for the SCO method w.r.t. $K$ to avoid an overwhelming number of lines.

A similar behavior is observed in Fig.~\ref{fig:3}, where we depict the fair optimization ($\omega_i \propto 1/\rho_i$). However, now, when the number of elements $L$ grows, there is a notable gap between the SCO and the benchmarking schemes. More precisely, our proposal almost doubles the AO performance, thus justifying the adequacy of SCO. An enhancement is also seen w.r.t. the $K$-sweep, further underscoring the potential of RIS. Unsurprisingly, the RO design always yields the poorest WSR values.

\subsection{Complexity Analysis}
Regarding the computational complexity of both methods, in Fig.~\ref{fig:4} we first evaluate the convergence tolerance (i.e., the relative difference between current and final rates) versus the number of iterations for $L = 16$ reflecting elements. The SCO procedure generally requires fewer steps to find the ultimate solution. Besides, according to the average execution times (per iteration) provided in Table~\ref{tab:1}, one can safely state that SCO surpasses AO (brute force quickly becomes infeasible). 

Notably, due to the convex formulation, the complexity of the SCO technique is indeed polynomial in $L^2$ (the number of design variables) \cite{Boy04}, whereas that of the AO algorithm grows exponentially with the search space. This can be verified in the average iteration times shown in Table~\ref{tab:1}. In both cases, though, the cost is independent of $K$ and $M$.

\section{Conclusions} \label{sec:5}
In this letter, we have addressed the problem of maximizing the WSR under the FBLR in an RIS-aided mMTC+ network with I-CSI errors. When considering a scenario with single-antenna sensors and a multiple-antenna CN, we have derived iterative procedures based on SCO to solve the nonconvex WSR problem and find suboptimal but feasible designs of the RIS reflection matrix. Numerical evaluations emphasize the performance of SCO, especially for large surfaces and arrays.

\begin{table}[t]
\caption{Average iteration time (in ms) w.r.t. RIS size $L$.}
\begin{center}
\begin{tabular}{|c|c|c|c|c|}
\hline
$L$ & AO (equal) & AO (fair) & SCO (equal) & SCO (fair) \\ 
\hline
$9$ & 8602.26 & 8641.63 & 25.83 & 19.42 \\
\hline
$16$ & 19924.69 & 19916.22 & 43.89 & 47.97 \\
\hline
$25$ & 222552.28 & 219705.37 & 133.21 & 141.49 \\
\hline
\end{tabular}
\label{tab:1}
\end{center}
\end{table}

\section*{Appendix} \label{app}
To derive the SINR with MRC receive processing and I-CSI, let us consider the use of (orthogonal) training sequences of length $N \geq M$ plus the MMSE criterion for estimating the channel matrix $\mathbf{H}_i$. For simplicity, we assume its elements are obtained sequentially via the on/off method \cite{Wei21}. Accordingly, for a coherence time of $T$ symbols, this algorithm needs $T \gg LN$ (although the required temporal constraint can be relaxed via grouping techniques or compressed sensing \cite{Guo20}).

Let the vectorized form of the linear MMSE estimate of the cascaded link $\mathbf{H}_i$ be $\hat{\mathbf{h}}_i \triangleq \textrm{vec}(\hat{\mathbf{H}}_i)  = \mathbf{A}_i \mathbf{z}_i + \mathbf{b}_i$, where $\mathbf{z}_i \in \mathbb{C}^{KL}$ is the (vectorized) observation obtained from pilot correlation \cite{Elw23}. The terms $\mathbf{A}_i$ and $\mathbf{b}_i$ follow from the statistics of the cascaded channel:
\begin{equation}
    \mathbf{A}_i = \mathbf{C}_i (\mathbf{C}_i + \sigma_i^2 \mathbf{I}_{KL})^{-1}, \quad \mathbf{b}_i = (\mathbf{I}_{KL} - \mathbf{A}_i)\bm{\mu}_i,
    \label{eq:23} 
\end{equation}
with $\bm{\mu}_i$ and $\mathbf{C}_i$ the mean and covariance of the (vectorized) cascaded channel $\mathbf{h}_i \triangleq \textrm{vec}(\mathbf{H}_i)$, respectively. In writing \eqref{eq:23}, we have defined $\sigma_i^2 \triangleq \sigma_w^2/(N P_i)$ for brevity.

Following the approach taken in \cite{Has03}, when conditioning on the set of channel observations $\mathbf{z}_i$, the data rate in \eqref{eq:5} can be lower bounded by $R_i(\rho_i)$ with equivalent SINR given by
\begin{equation}
    \rho_i = \frac{P_i \vert \mathbf{f}_i^{\textrm{H}} \hat{\mathbf{H}}_i \bm{\psi} \vert^2}{\sigma_w^2 \| \mathbf{f}_i \|_2^2 +  \textrm{tr}(\mathbf{C}_e(\bm{\psi} \bm{\psi}^{\textrm{H}} \otimes \mathbf{f}_i\mathbf{f}_i^{\textrm{H}})) + \sum_{j \neq i} P_j  \vert \mathbf{f}_i^{\textrm{H}} \hat{\mathbf{H}}_j \bm{\psi} \vert^2},
\label{eq:24}
\end{equation}
where $\mathbf{C}_e \triangleq \sum\nolimits_{j=1}^M P_j (\mathbf{C}_j - \mathbf{A}_j \mathbf{C}_j)$ is the sum of covariance matrices of the estimation errors $\mathbf{h}_i - \hat{\mathbf{h}}_i$.

After some manipulations, the covariance matrix can be formulated as $\mathbf{C}_i = \mathbf{I}_L \otimes \bar{\mathbf{C}}_i$ (cf. \eqref{eq:2} and \eqref{eq:3}), where
\begin{equation}
    \bar{\mathbf{C}}_i \triangleq \alpha \beta_i/((1 + \gamma)(1 + \delta_i))((1 + \delta_i)\mathbf{I}_K + \gamma \ddot{\mathbf{G}} \ddot{\mathbf{G}}^{\textrm{H}}),
\label{eq:25}
\end{equation}
which holds for the near- and far-field regions.

Based on that, the (total) estimation error matrix can also be decomposed as $\mathbf{C}_e = \mathbf{I}_L \otimes \tilde{\mathbf{C}}$, with
\begin{equation}
    \tilde{\mathbf{C}} \triangleq \sum\nolimits_{j = 1}^M  P_j (\bar{\mathbf{C}}_j - \bar{\mathbf{C}}_j (\bar{\mathbf{C}}_j + \sigma_j^2 \mathbf{I}_{K})^{-1} \bar{\mathbf{C}}_j),
\label{eq:26}
\end{equation}
which allows us to rewrite the second term in the denominator in \eqref{eq:24} as $\| \bm{\psi} \|_2^2 \mathbf{f}_i^{\textrm{H}}\tilde{\mathbf{C}} \mathbf{f}_i$. Finally, when substituting the MRC beamformers $\mathbf{f}_i = \hat{\mathbf{H}}_i \bm{\psi}$ in the aforementioned SINR $\rho_i$, one can obtain the expression in \eqref{eq:6}. This concludes the proof.

\bibliographystyle{IEEEtran}
\bibliography{./references}

\end{document}